\documentclass[reprint,superscriptaddress,footinbib,aps,pre,showkeys]{revtex4-1}
\usepackage{subfigure}
\usepackage{graphicx}
\usepackage{citesort}
\usepackage[letterpaper,centering,margin=1.8cm]{geometry}

\begin{document}

\title[Protein multiscale organization through graph partitioning and robustness analysis]{Protein multi-scale organization through graph partitioning and robustness analysis: Application to the myosin-myosin light chain interaction}

% \author{A.~Delmotte$^{1,3,\dagger}$}
% \email{antoine.delmotte09@imperial.ac.uk}
% \address{Department of Mathematics, $^2$Department of Chemistry, $^3$British Heart Foundation Centre of Research Excellence, Imperial College London, South Kensington Campus, London SW7 2AZ, UK}
% \author{E.W.~Tate$^{2,3}$}
% \author{S.N.~Yaliraki$^{2,3}$}
% \author{M.~Barahona$^{1,3,\ast}$}
% \email{m.barahona@imperial.ac.uk}
\affiliation{Department of Mathematics, Imperial College London, London SW7 2AZ, UK}
\affiliation{Department of Chemistry, Imperial College London, London SW7 2AZ, UK}
\affiliation{British Heart Foundation Centre of Research Excellence, Imperial College London, London SW7 2AZ, UK}
%\affiliation{Imperial College London, South Kensington Campus, London SW7 2AZ, UK}

\author{A.~Delmotte}
\email{antoine.delmotte09@imperial.ac.uk}
\affiliation{Department of Mathematics, Imperial College London, London SW7 2AZ, UK}
\affiliation{British Heart Foundation Centre of Research Excellence, Imperial College London, London SW7 2AZ, UK}
%\affiliation{Imperial College London, South Kensington Campus, London SW7 2AZ, UK}

\author{E.W.~Tate}
\author{S.N.~Yaliraki}
\affiliation{Department of Chemistry, Imperial College London, London SW7 2AZ, UK}
\affiliation{British Heart Foundation Centre of Research Excellence, Imperial College London, London SW7 2AZ, UK}
%\affiliation{Imperial College London, South Kensington Campus, London SW7 2AZ, UK}

\author{M.~Barahona}
\email{m.barahona@imperial.ac.uk}
\affiliation{Department of Mathematics, Imperial College London, London SW7 2AZ, UK}
\affiliation{British Heart Foundation Centre of Research Excellence, Imperial College London, London SW7 2AZ, UK}
%\affiliation{Imperial College London, South Kensington Campus, London SW7 2AZ, UK}

%\email{$^\dagger$antoine.delmotte09@imperial.ac.uk, $^\ast$m.barahona@imperial.ac.uk}

\begin{abstract}

Despite the recognized importance of the multi-scale spatio-temporal organization of proteins, most computational tools can only access a limited spectrum of time and spatial scales, thereby ignoring the effects on protein behavior of the intricate coupling between the different scales. 
Starting from a physico-chemical atomistic network of interactions that encodes the structure of the protein, we introduce a methodology based on multi-scale graph partitioning that can uncover partitions and levels of organization of proteins that span the whole range of scales, revealing biological features occurring at different levels of organization and tracking their effect across scales. 
Additionally, we introduce a measure of robustness to quantify the relevance of the partitions through the generation of biochemically-motivated surrogate random graph models. We apply the method to four distinct conformations of myosin tail interacting protein, a protein from the molecular motor of the malaria parasite, and study properties that have been experimentally addressed such as the closing mechanism, the presence of conserved clusters, and the identification through computational mutational analysis of key residues for binding.

\end{abstract}


%\submitto{\PB}
\maketitle

\section{Introduction}
\subsection{The methodology: Multi-scale analysis of protein structures}
Proteins are complex structures characterized by multiple scales in time and space~\cite{kern,kernAdK,Frauenfelder2001,mauriciosophia}. Atoms, functional chemical groups, amino acids, the ensuing secondary structures, the large conformational domains: each define different, yet coupled, levels of structural and dynamical organization linked to behaviors occurring at different time and spatial scales. Molecular dynamics simulations~\cite{Adcock2006} can deal successfully with the very short time scales but such methods cannot be applied to long times (or large systems) due to their exorbitant computational cost. On the other hand, because many of the key biological functions take place at the micro- to millisecond time scales, strongly simplified coarse-grained systems have been proposed as a means to reaching the biologically relevant regimes~\cite{NMArev1,Tozzini,Ayton2007,los_rios_HNM}. However, these simplified coarse-grainings often ignore the detailed physico-chemical atomic interactions and, consequently, cannot provide a picture that emerges seamlessly from the smallest scales. 

Yet the different levels of organization in proteins do not behave independently: the dynamics at long time and length scales, which is in many cases crucial for biological function, is the result of the integrative interaction of the finer organizational levels. Analyzing proteins from this multi-scale perspective can reveal the intricate linkage between the levels and give insight into the behavior of the protein starting from the bottom-up. This picture can also aid in the understanding of the effects that small-scale changes like mutations have on large-scale organization. To achieve this, we apply a recently proposed general methodology for multi-scale graph partitioning~\cite{JCPNAS} that uses dynamical processes on graphs to uncover the multi-level organization of graph communities relevant at different time scales. 
This method allows us to bridge the gap across scales, thus relating the behavior at a certain level to the consequences it has at coarser scales. In the case of proteins, our analysis starts from a fully atomistic description of the protein which is transformed into a graph theoretical formalism. The method is then able to find increasingly large clusters of atoms that behave coherently over increasingly long time scales and quantifies the time scales over which those groupings are relevant. This leads to a multi-level hierarchical organization of the protein structure at different scales: from chemical groups through amino acids, to the appearance of secondary structures and intermediate structural elements, such as clusters of several helices, to the eventual emergence of large conformational units~\cite{Stefano}. Hence the picture at larger scales emerges directly from the detailed physico-chemical information at the smallest atomic scales. 

In this paper, we extend this multi-scale methodology and then apply it to understand the multi-scale dynamical features of a class of biologically relevant proteins and to infer possible mechanisms of functional motions. The present work extends the methodology in two ways: firstly, we introduce a tool to quantify the relevance of a level of organization through a novel measure of its robustness as compared to that of relevant biochemically-motivated surrogate null models;
secondly, we introduce a measure which estimates the effect of mutations of a particular residue on the structure and dynamics of the global properties of the protein and therefore suggests key residues or ``hotspots'' that could be targeted through mutagenesis.
This extended method is first exemplified on adenylate kinase (AdK), a model protein for which there is extensive experimental data. We then apply our analysis to the detailed study of a particular myosin-myosin light chain interaction, an example of biological importance in which protein interactions lead to significant changes in their functionality. We now give some biological background for this biological system.

\subsection{The biological system: Myosin-myosin light chain interaction}
\begin{figure}
 \centering
 \includegraphics[width=8.2 cm]{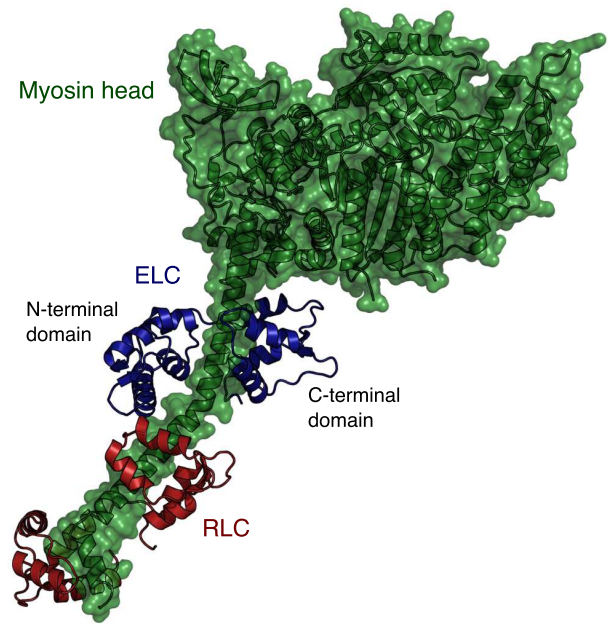}
 % figure1.eps: 0x0 pixel, 300dpi, 0.00x0.00 cm, bb=0 -1 451 388
 \caption{Crystal structure of the scallop muscle myosin essential (ELC) and regulatory (RLC) light chains in complex with the myosin heavy chain (PDB ID: 1QVI). This ELC is the closest structural MTIP homolog \cite{Bosh2006} and, when bound to the myosin heavy chain, adopts a conformation similar to the way MTIP wraps around the MyoA tail.}
 \label{fig:figure1}
\end{figure}

Myosin light chains (MLCs) are small proteins known to play a major role in the regulation of motor complexes. In mammalian muscle cells, the essential and regulatory light chains regulate the actin-myosin motor complex by binding to the myosin heavy chains. Each of the myosin heavy chains consists of a long tail terminated by a globular head, where the actin binds upon activation by ATP. Preceding the head is the neck region, formed by a single $\alpha$-helix, which serves as the binding domain for the two light chains (Fig. \ref{fig:figure1}) \cite{Lowey2010,schiaffino96}. Crystal structures of the myosin head and neck regions \cite{rayment1993} have highlighted the importance of the light chains in stabilizing the lever arm formed by these two regions, which allows a more powerful stroke in the cross-bridge cycle \cite{Yamashita2003}. The light chains have also been suggested to be responsible for the fine tuning of the motor apparatus and even to interact directly with the actin filaments \cite{Timson2003}. However, the structure and dynamics of the myosin light chains and their effects on regulation are still poorly understood.

Here we focus on myosin tail interacting protein (MTIP), a myosin light chain involved in the invasion machinery of \textit{Plasmodium} species, which include the causative agents of malaria. Therefore, MTIP is of particular interest as a potential target for the design of anti-malarial drugs~\cite{Jemima2010}. MTIP binds myosin A (MyoA), an unconventional class XIV myosin, which was first found to be part of the motor complex responsible for the gliding mobility of \textit{Toxoplasma gondii} \cite{Heintzelman1997}. Later on, MyoA was also identified in \textit{Plasmodium} species and discovered to be responsible for their ability to invaginate red blood cells \cite{Hettmann2000}. MyoA was then found to be anchored to an inner membrane complex, located just behind the plasma membrane, via the MTIP protein \cite{Bergman2002,Green2008,ReesChanner2006,Green2006,Baum2006,Frenal2010}. 

A first crystal structure containing three conformations of \textit{Plasmodium knowlesi} MTIP (PkMTIP) suggested that its binding to MyoA should essentially be realized by the two lobes of the C-terminal domain wrapping around the MyoA tail \cite{Bosh2006}. However, subsequent crystal structures of \textit{Plasmodium falciparium} MTIP (PfMTIP), and binding assays demonstrated the importance of the N-terminal domain, suggesting that it should also change its conformation to bind with the MyoA tail \cite{Jemima2010,Bosch2007}. Although initial assays suggested the last fifteen residues to be responsible for most of the interactions \cite{Bergman2002}, subsequent studies showed that the last nineteen residues give a much stronger binding \cite{Jemima2010}.

To help resolve these discrepancies, we study here how different parts of MTIP interact dynamically at different time scales, and how these interactions are influenced upon binding with the MyoA tail. We investigate how the mechanism by which MTIP wraps around the MyoA tail is related to the multi-scale structure of the protein. Our goal is to understand the changes induced in the structure upon binding with the MyoA and to identify amino acids of the MyoA tail that play a key role in the binding.

The paper is organized as follows. Firstly, we describe our multi-scale graph partitioning algorithm and introduce two novel biochemically-motivated random graph models, which allow our analysis to target a specific level of organization by including in the null model biochemical properties dominant at the different scales considered. Secondly, the method is tested on \textit{Escherichia coli} adenylate kinase, since its structure and closing mechanism are well documented~\cite{kernAdK}. Thirdly, the dynamic behavior of MTIP is investigated by identifying parts of the protein sharing the same dynamics at a particular scale, leading to a hypothetical closing mechanism. The same tool is then used to explain the differences observed between different conformations of PkMTIP and between the structures of PkMTIP and PfMTP. Finally, the role of each amino acid of the MyoA tail is probed through computational mutagenesis.

\section{Materials and methods}
\label{sec:methods}

\subsection{Structural data}

We first apply our extended methodology to a crystal structure of \textit{Escherichia coli} AdK (PDB ID code 4AKE). We then study in detail four crystal structures of \textit{Plasmodium knowlesi} (Pk) and \textit{Plasmodium falciparium} (Pf) MTIPs, either unliganded or in complex with a MyoA tail peptide. The two unliganded PkMTIP and the complexed PkMTIP/MyoA structures were obtained by Bosch \emph{et al} \cite{Bosh2006} through 2.6 \AA\ resolution X-ray crystallography and comprise residues K79 to L204 of \textit{P. knowlesi} MTIP and residues S803 to A817 of \textit{P. yoelii} MyoA tail solved at pH 5.3 (PDB ID code 2AUC). The three conformations were found within the same asymmetric unit in the crystal. The structure of PfMTIP in complex with a 15-amino acid MyoA tail peptide, which has been determined by the same group \cite{Bosch2007}, is a 1.7 \AA\ resolution crystal structure comprising residues E60 to Q204 in complex with the same \textit{P. yoelii} MyoA tail peptide and solved at pH 7.5 (PDB ID code 2QAC).

\subsection{Multi-scale graph partitioning through Stability}

When dealing with complex graphs, it is sometimes desirable to obtain simplified reduced representations in terms of subgraphs or communities, i.e., meaningful groupings of nodes that are significantly related. For instance, the nodes of a network are likely to belong together if they are part of a tightly-knit group with many connections within the group and fewer to external nodes. Such communities can then be used as coarse-grained representations of the network~\cite{Delosrios_PRL}. Community detection and graph clustering has a long history, and recent research has both rediscovered classic results and introduced novel methods~\cite{fortunato_review,Gfeller2007}. 

In this work, we use Stability, a recently introduced method for multi-scale graph partitioning~\cite{JCPNAS,renaud} that is particularly suited to the analysis of structures, such as proteins, with an intrinsic multi-scale organization. Stability uses a dynamical (Markov) process taking place on the graph to establish its community structure. A community is relevant over a particular time scale if the dynamical process tends to be more contained inside that group over that time scale than would otherwise be expected at stationarity. Hence, our method has an intrinsic Markov time associated with the dynamics that reveals the community structure at different scales. The analysis can be viewed as following the time evolution of a linear probabilistic process on the graph and identifying the subgraphs where the probabilistic flow 
gets trapped. This is measured in terms of the Stability $R(t)$, which can be seen as a clustered autocovariance:
\begin{equation} \label{2stab}
R(t)=\sum_{C} \sum_{i,j \in C} \left[\left(e^{- t \, L /\langle k \rangle}\right)_{i,j} \frac{1}{N} - \frac{1}{N^2}\right], 
\end{equation}
where $C$ extends over the set of communities and $i,j$ extend over the $N$ nodes of the graph. Each node of the graph
has a degree $k_i$. Here $\langle k \rangle$ is the average degree, and $L$ is the Laplacian matrix: $L= \mathrm{diag}(k_i) - A$, where $A$ is the adjacency matrix of the graph.
Hence, as the Markov time increases, Stability follows the expanding transient of this dynamics towards stationarity and, in doing so, it allows us to reveal naturally a sequence of coarser partitions that uncovers the multi-scale structure of the graph, if it exists.

\subsubsection{Application of multi-scale Stability partitioning to proteins. }

In our original work, we already indicated how this generic methodology for graph analysis provides a route for the analysis of the multi-scale organization of protein structures~\cite{JCPNAS}. Subsequently, the method has been refined and tested extensively on a variety of protein structures encoded in terms of a weighted graph formalism that is built bottom-up from the atomistic description of the protein~\cite{Stefano}. In this abstraction, each atom is represented by a node, and each covalent bond or weak interaction (hydrogen bonds, hydrophobic tethers and salt bridges) by an edge associated with a weight related to the potential energy of that interaction.

The graph is generated as follows. We start from a PDB file~\cite{pdb} containing the spatial coordinates of each atom of the molecule and relax it through energy minimization using the molecular dynamics package GROMACS~\cite{gromacs}.
When necessary, GROMACS is first used to add missing hydrogens. The protein is then placed in a cubic box of spc216 water separated from the walls of the box by a distance of 0.6 nm, and an energy minimization of the structure is carried out using the steepest descent and conjugate gradient algorithms included in GROMACS. Once the structure is relaxed, we use the software FIRST~\cite{FIRST} to identify the bonds and interactions present in the network. These constitute the edges of the graph. The interaction potentials are here approximated by mass-spring systems, with a specific spring constant that defines the weight of the corresponding edge~\cite{Stefano,stefanopreprint}. 
Ours is a distinct variation from other approaches to generate network representations of proteins~\cite{FIRST,costa,Chennubhotla2006,Bahar1997}, in that most of those other approaches generate unweighted graphs or graphs based on proximity edges and often coarse-grain to the level of amino acids (see also \cite{bahar2010} and references therein). Our weighted network of bonds and weak interactions includes full details of the chemistry as well as information about the spatial conformation since the location of edges partly encodes the relative position of atoms and has proved to be an efficient description for the study of biomolecules.

The structural and dynamical organization of the protein is then extracted by identifying the communities (i.e., groups of atoms) 
that are relevant at different time scales according to the Stability~(\ref{2stab}) which, importantly, applies directly to weighted graphs. Previous work~\cite{JCPNAS,Stefano} has shown that this approach is able to identify meaningful partitions at different scales, from bonds and chemical groups to large functional domains, by sweeping the intrinsic Markov time of the algorithm. The Markov time can be related monotonically to the biophysical time of motion of the corresponding groups of atoms as compared to experiment or atomistic simulation~\cite{Stefano,stefanopreprint}. Hence the multi-scale groupings found through Stability establish a link between substructures at particular spatial scales and dynamics at time scales specified by the Markov time. In particular, we then establish computationally the effect of mutations on this multiscale organization. 

Chennubhotla and Bahar~\cite{Chennubhotla2006} studied allostery in the complex GroEL-GroES using a coarse-graining algorithm that preserves properties of the stationary distribution of a Markov process on the network of residues. Sehti \emph{et al.}~\cite{Sethi2008} constructed a network based on correlations of fluctuations in 20 ns molecular dynamics simulations and used graph partitioning based on another quality function, the modularity, to identify residues critical for allostery. However, our work differs from these methods in two important aspects: Firstly, the network in those methods is coarse-grained at the level of residues, whereas in our method amino acids emerge as a natural partition from the network of atomistic interactions. Secondly, the distinctive feature of our method (i.e., the uncovering of the multi-scale spatial structure relevant over different time scales) is  not present in any of those methods.

\subsubsection{Algorithm for the optimization of Stability. }

Stability, as defined in~(\ref{2stab}), provides a measure to assess the quality of a defined partition. However, the global optimization of Stability is computationally hard---a common occurrence in the study of complex landscapes. We can use a variety of heuristic algorithms to obtain good partitions which can then be ranked by Stability to provide us with near-optimal partitions at different time scales.
Different such algorithms exist, either greedy or divisive, depending on whether nodes are progressively grouped to form communities, or whether the whole graph is gradually divided into smaller groups of nodes. 

Here, we use a greedy agglomerative method, the Louvain algorithm \cite{louvain,louvain2}, which has been shown to provide an extremely efficient optimization of Stability. Briefly, Louvain works as follows. Initially, each node is assigned to its own community. The 
nodes are successively transferred into the neighboring community where the increase of Stability is the biggest, as long as it improves the Stability of the overall partition. This step is repeated until no transfer can increase the stability. At that point, a new meta-graph of communities is generated, and the algorithm repeats these two steps until a graph is obtained where no further grouping can improve the stability. This heuristic has been observed to require little computational effort and to find partitions close to the optimal solution.
Note that the method is deterministic but the final solution found depends on the order in which the different nodes are scanned for the grouping step. This can be chosen at random every time the algorithm is run, and we will refer to it in what follows as the \textit{Louvain initial condition}. Indeed, we will use the variability of the observed solution induced by our random choice of the Louvain initial condition to estimate the robustness of a partition, a measure of its relevance.

\subsection{Robustness tools for Stability analysis}
At each Markov time, a different partition with optimal Stability can potentially be obtained. However, not all optimal partitions are meaningful. Therefore, the question that now needs to be addressed is: which partitions, among all those generated across Markov times, are relevant? In this paper, we introduce robustness tools to address this issue, which is of general importance in multi-scale analysis methods, and we provide specific robustness tools for the analysis of proteins.

\subsubsection{Identification of relevant partitions and robustness analysis. } \label{sec:robustness}
As suggested by Karrer~\emph{et al}~\cite{karrer}, the defining property of a significant community structure should be its robustness with regard to small perturbations. A partition is robust if, when introducing an alteration, either of the graph itself or of the partitioning method, the new partition found by any method is very similar to the one obtained originally. In this sense, the ``Markov lifetime'' of a partition, i.e., how long the partition is optimal in terms of Stability, is a straightforward way to obtain an initial assessment of its robustness and relevance~\cite{Stefano,Fenn}.

An alternative way of measuring the robustness, and thus the significance, of a partition consists in quantifying to what extent the result is changed by a perturbation~\cite{karrer,MRes_project1,montjoye}. This can be done by measuring the distance between the solutions found before and after the perturbation. The distance between two partitions can be measured by the \textit{variation of information}, a true metric of the amount of information not shared by two partitions~\cite{meila}. 
Consider a community $C_k$ containing $n_k$ nodes among the $N$ of the whole network. Let $f_k = \frac{n_k}{N}$ be the fraction of the nodes belonging to community $C_k$. The amount of information contained in a partition $\mathcal{P}$ can then be defined by its Shannon entropy
\begin{equation}
\mathrm{H}(\mathcal{P})= - \sum_{k} f_k \log f_k.
\end{equation}
The variation of information (VI) between two partitions $\mathcal{P}$ and $\mathcal{P'}$, relating to how much information is not shared by $\mathcal{P}$ and $\mathcal{P'}$, can be expressed as a function of their marginal ($\mathrm{H}(\mathcal{P})$, $\mathrm{H}(\mathcal{P'})$) and joint ($\mathrm{H}(\mathcal{P,P'})$) entropies
\begin{equation}
\mathrm{VI}(\mathcal{P}, \mathcal{P'}) = 2 \mathrm{H}(\mathcal{P}, \mathcal{P'}) - \mathrm{H}(\mathcal{P}) - \mathrm{H}(\mathcal{P'}),
\end{equation}
where 
\begin{equation}
\mathrm{H}(\mathcal{P,P'})= - \sum_{k} \sum_{k'} f_{k,k'} \log f_{k,k'}.
\end{equation}
% 
% The probability of a node to be a member of $C_k$ is $P(k) = n_k/N$. This defines a discrete random variable $\mathcal{P}$ which can take $N_C$ discrete values, $N_C$ being the number of clusters. Each node then corresponds to a different event of $\mathcal{P}$, and the information of a partition is simply defined by the Shannon entropy,
% \begin{equation}
% \mathrm{H}(\mathcal{P})= - \sum_{k} P(k) \log P(k)
% \end{equation}
% The variation of information (VI) between two partitions $\mathcal{P}$ and $\mathcal{P'}$ relates to how much information is not shared by the two random variables associated with each clustering
% \begin{equation}
% \mathrm{VI}(\mathcal{P}, \mathcal{P'}) = \mathrm{H}(\mathcal{P})+\mathrm{H}(\mathcal{P'}) - 2 \mathrm{I}(\mathcal{P},\mathcal{P'}),
% \end{equation}
% where $\mathrm{I}(\mathcal{P},\mathcal{P'})$ is the mutual information, namely the information shared by the two partitions, defined by
% \begin{equation}
% \mathrm{I}(\mathcal{P},\mathcal{P'}) = \sum_{k} \sum_{k'} P(k,k') \log \frac{P(k,k')}{P(k)P(k')}.
% \end{equation}
As defined, this measure depends on the size of the network: larger networks contain more information. Therefore, in this work, we use a normalized version of the VI by dividing it by its maximum, $\log N$.

As stated above, the perturbation used here consists in changing the Louvain initial condition. By computing the partitions with 100 different initial conditions, the distances between all pairs of solutions are calculated, and the average is used as a measure of how much the partitions are affected by the perturbation, which we then use as an estimate of their robustness.

\subsubsection{Surrogate random graph models. } \label{rdg}

The normalized variation of information does not, in itself, give an absolute value of the robustness of the partitions since the number of possible partitions varies with the number of communities found, which changes with the Markov time. 
This problem can be overcome by comparing the VI at each Markov time against a surrogate control group, obtained from 
a random graph model. The use of random graph surrogate models is a classical bootstrapping tool in graph theory~\cite{newmanbook} to test the emergence of particular statistical properties in a certain type of graph, to classify graphs into different categories, or to highlight differences in the properties of different types of graphs. 
Here we use the z-score statistic to compare the robustness of the partitions of a particular graph with an ensemble of graphs from the random graph model, as follows. For each Markov time $t$, generate $K$ surrogate graphs from the random model. For each of those $K$ graphs, obtain the average VI($t$) computed between all pairs of partitions obtained by starting from 100 Louvain initial conditions. Then compute the mean $\mu(t)$ and standard deviation $\sigma(t)$ over the $K$ average VI values of the surrogates. The z-score of the variation of information then reads:
\begin{equation}
 Z(t) = \frac{VI(t) - \mu(t)}{\sigma(t)},
\end{equation}
which we can then use as an estimate of the robustness of the partition independent from the number of communities detected. 
%It will also be used to compare the robustness of partitions with a same number of communities, in which case it becomes:
%\begin{equation}
% Z(M) = \frac{VI(M) - \mu(M)}{\sigma(M)}
%\end{equation}
%where $M$ is the number of communities.

%%%%%%%%%%%%%%%%  
%%%%  Results

\section{Results and discussion}

\subsection{Biochemical null models to estimate the robustness of protein partitions at different scales}

\begin{figure}[htp]
  \centering
  \subfigure[]{
  \includegraphics[width=7cm]{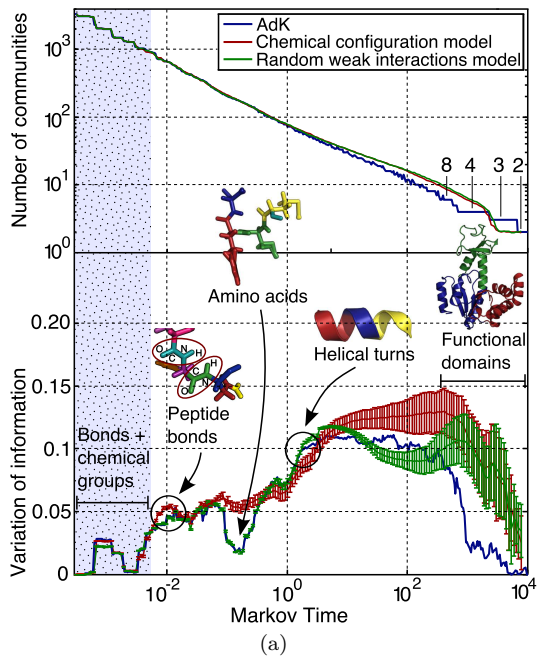}
  \label{fig:AdKa}
  }\\
  \subfigure[]{
  \includegraphics[width=6.5cm]{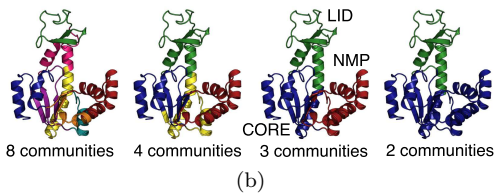}
  \label{fig:AdKc}
  }\\
  \subfigure[]{
  \includegraphics[width=4.5cm]{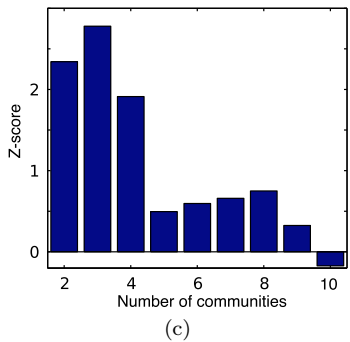}
  \label{fig:AdKb}
  }
\caption{(a) Multi-scale partitioning of AdK as a function of the Markov time. The comparison of the robustness of the protein structure with the two types of biological random graph surrogates allows us to detect partitions which are biochemically meaningful at different scales, such as peptide bonds, amino acids, the emergence of single helical turns and the appearance of functional domains. The shaded area on the left corresponds to the Markov times for which both null models and the protein give the same partitions due to the fact that the local chemistry is preserved. Error bars represent the standard deviation of the optimal partitions from the different random graphs. (b) Relevant partitions of AdK at large Markov times. The partition into 8 communities relates to previous hinge analyses~\cite{kernAdK,Stefano}, while the 3-way partition corresponds to the functional domains. (c) Using the z-score to compare partitions with the same number of communities between AdK and the ensemble of graphs with randomized weak interactions, the partition into three communities, which divides AdK into its functional domains, is indeed identified as the most meaningful. Protein structures drawn with PyMOL~\cite{PyMOL}.}\label{fig:AdK}
\end{figure}

We introduce our extended method through the analysis of an example that has been well studied both experimentally and computationally~\cite{kernAdK}, namely the adenylate kinase (AdK) from \textit{Escherichia coli}. The analysis proceeds as described in section~\ref{sec:methods}. We start from the corresponding PDB file, relax the structure through energy minimization, and then obtain a weighted graph representation with edges based on identifying physico-chemical interactions. We then find partitions that optimize Stability at different Markov times. In Fig.~\ref{fig:AdK} we show that as the Markov time increases, the optimal partition gets coarser: at very small values, each atom is identified as a distinct community; at very large times, the graph is partitioned into two large communities. Both at low times and large times, it is apparent that certain partitions have long persistence, i.e., they remain optimal over long intervals of the Markov time. This persistence is an indication of their relevance at the corresponding time scales. 

However, it is difficult to establish the persistence of partitions in the intermediate regime of the Markov time. 
This is partly due to the fact that the number of possible partitions of intermediate size grows combinatorially. In order to refine the evaluation of the robustness of the partitions, we calculate, at each Markov time, the variation of information (VI) between 100 optimal solutions found starting from 100 random Louvain initial conditions and compare it with the VI of surrogate random graph models using a z-score statistic. The ensemble of surrogate models can be designed to test the null hypothesis. 

In this particular case, we use our intrinsic structural knowledge of the physico-chemical structure of proteins to formulate surrogates that can probe the emergence of biochemically relevant substructures at different scales. Indeed, the multi-scale organization observed in the case of proteins is particularly interesting because communities at different levels are linked to the presence of edges of different biophysical origin. For instance, the organization of the protein in the form of a chain of amino acids is only defined by the network of covalent bonds, while higher levels of organization, such as conformational and tertiary structures, only depend on the position of the weak interactions and are essentially independent from the organization of the covalent bonds. The biophysical origin of the different forms of structural organizations, which can either be chemical or spatial, leads to the definition of two types of surrogate random graph models for the robustness analysis.

\paragraph{Robustness analysis at short scales: the chemical configuration model. }
Our first surrogate set is based on a random graph that preserves the local chemistry of the protein while randomizing all other interactions. This can be used as a chemical null model that should be identical to our original graph at short time and length scales but will highlight the differences that emerge with the longer scale organization. The random graph model is designed to preserve the chemical attributes of the protein including the chemical composition of the molecule preserving the valence of the atoms, encoded in the degree of the nodes, and the energies of the bonds and interactions, encoded in the weights of the edges. 
All the basic chemical properties of the graph can be kept using a simple randomization scheme similar to the one proposed by Maslov and Sneppen~\cite{swap}, in which pairs of bonds chosen at random exchange one of the two nodes they link. 
By doing this repeatedly, a new random graph keeping the number but also the weights of the connections of each node is generated. The same method is used here, with the additional constraints that the pairs of bonds are of the same kind (covalent bonds of the same energy, or weak interactions of the same nature), and that the exchange keeps the whole network of covalent bonds connected. This randomization thus also keeps the chemical nature of the neighbors of each atom. Consequently, from a chemical point of view, the small chemical groups are kept, and, from a graph theoretical point of view, the degree of each node is also maintained. In that respect, this model is similar to the configuration model~\cite{Molloy1995} and can be thought of as the ``chemical configuration model''. 

\paragraph{Robustness analysis at long scales: randomized weak interactions. } The large-scale spatial organization of the protein is mainly determined by the weak interactions such as hydrogen bonds, hydrophobic tethers or salt bridges. The second type of surrogate random graph therefore conserves the whole network of covalent bonds defining the primary structure of the protein, but randomizes the positions of the weak interactions which determine the secondary and tertiary structures. The randomization of these interactions is carried out preserving the necessary chemical constraints: hydrogen bonds should only bind oxygen or nitrogen with hydrogen atoms and hydrophobic tethers, carbon and sulphur atoms. The weak interactions are then re-positioned between nodes of the required nature selected at random.

\paragraph{Application to AdK. } 
Results of the partitioning and robustness analysis for AdK are summarized in Fig.~\ref{fig:AdK}. At each Markov time, the Stability
was optimized a hundred times with different Louvain initial conditions. For each Markov time, the optimal partition is shown
in the top panel of Fig.~\ref{fig:AdK}(a) by its number of communities, and the variation of information between all the partitions found at this Markov time is shown in the bottom panel.
Partitions at very small and very high Markov times remain optimal for extended Markov times and correspond to biochemically meaningful components: small chemical groups (small times) or the three functional domains (LID, NMP and CORE domains)~\cite{JCPNAS,Stefano,stefanopreprint}. This is confirmed by our robustness analysis, which shows small values of VI for the long-lived partitions. 

Fig.~\ref{fig:AdK} also shows the comparison of the robustness of the partitions of the protein against that of ensembles of 
random graphs from our surrogate models. As expected, the random graphs obtained from the chemical configuration model are indistinguishable from the protein at short Markov times, since their local chemical structure is identical. However, at longer times
the comparison reveals two additional partitions of strong biochemical significance corresponding to the peptide bonds between the amino acids (at Markov times around $10^{-2}$) and to the emergence of amino acids at Markov times around $10^{-1}$. At the local minimum of VI, 63\% of the amino acids were grouped as a community, while most of the others, essentially small amino acids, were grouped with another residue due to the slight tendency of Stability to find communities of about the same size.

The robustness of the protein is indistinguishable from the ensemble of graphs obtained by randomization of weak interactions
until Markov times of around 2. This establishes the spatial and time scale at which the weak interactions start having an influence on the communities, and, by extension, on the conformation of the protein in space. Interestingly, the communities found at this scale usually contain four amino acids, which is the number of residues usually found in one turn of an $\alpha$-helix.

At long Markov times, Stability finds partitions into a few subunits that are much more robust for the protein than for the random surrogates.
The study of their robustness indicates the relevance of partitions into 2, 3, 4, and 8 communities, which can be linked to the well-known folding of AdK and to an existent hinge analysis~\cite{Stefano,kernAdK}. As expected, the two random models converge at long Markov times since in both cases the composition of the molecule is conserved, and the weak interactions have been randomized. At Markov times above 30 we observe an increase in the variability and a decay in the value of the VI for the surrogates from the weak interaction randomization.
This indicates the point where weak interactions placed at random begin to induce robust compact subgroups in the structure in an effect akin to undirected packing. In contrast, the specific location of the weak interactions in the structure of the protein induces robust and reproducible partitions that reveal the specific organization of the protein conformation. 
	 
\subsection{The succession of partitions of PfMTIP suggests a rigid cluster and the dynamics of the closing mechanism} \label{sec:PfMTIP}

\begin{figure}
 \centering
 \includegraphics[width=8cm]{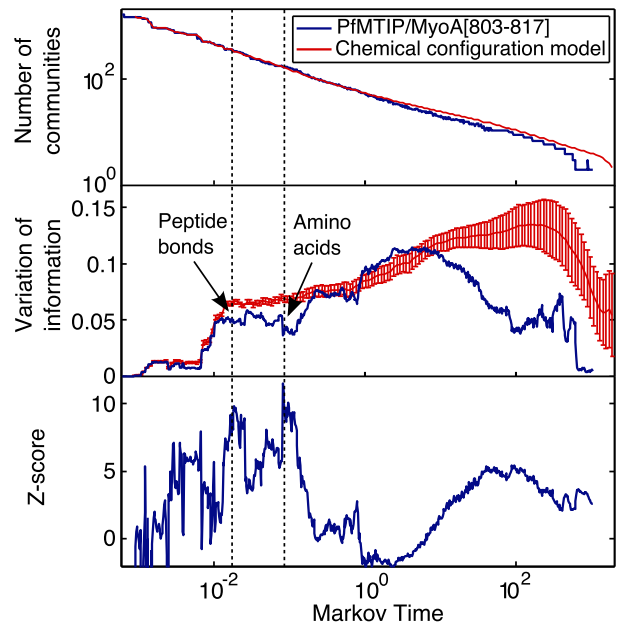}
 % PfMTIP_swap.eps: 0x0 pixel, 300dpi, 0.00x0.00 cm, bb=
 \caption{Using the z-score, the comparison across Markov times of the robustness of the partitions of PfMTIP and random graphs from the chemical configuration model identifies amino acids and peptide bonds as biochemically relevant communities.}
 \label{fig:PfMTIPswap}
\end{figure}

\begin{figure}[htp]
  \centering
  \subfigure[]{
  \includegraphics[width=7.6cm]{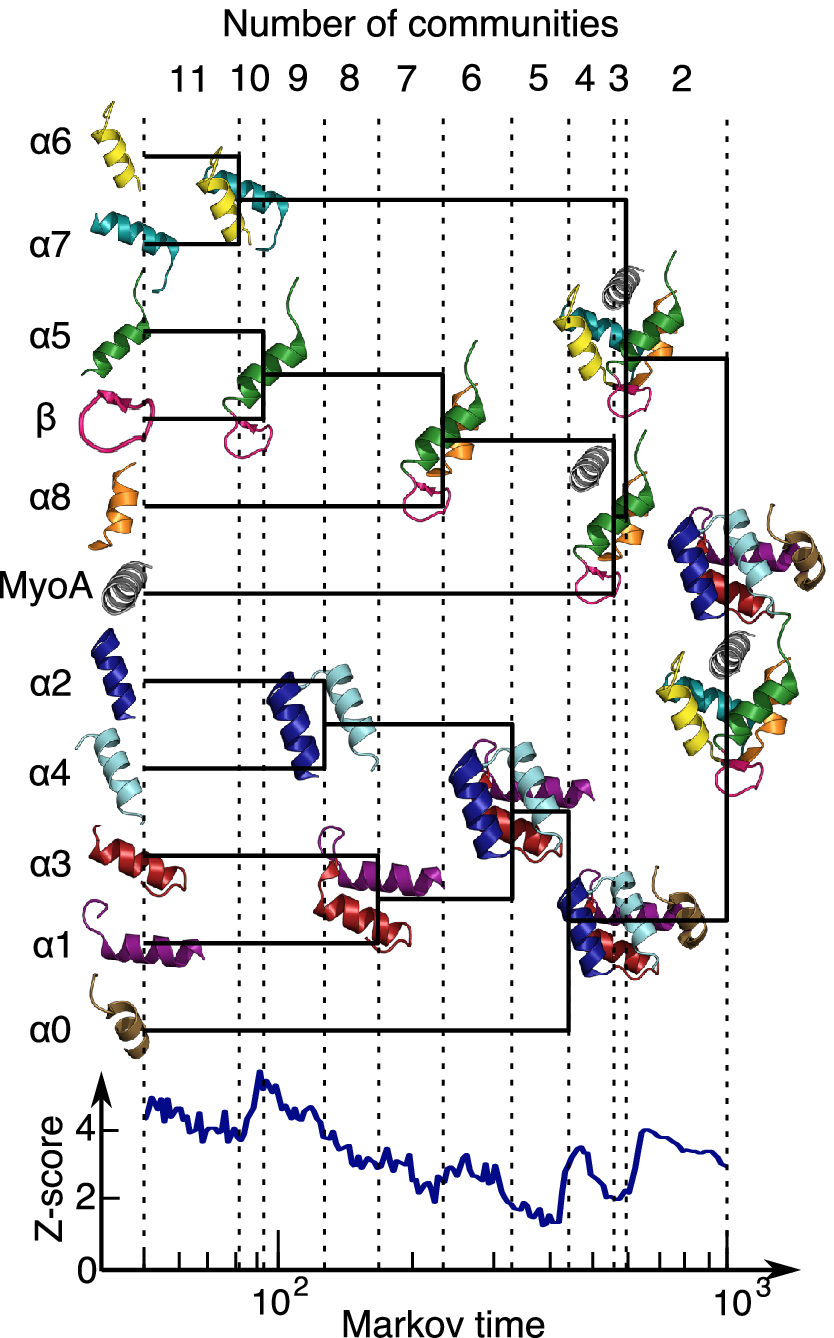}
  \label{fig:PfMTIPa}
  }\\
  \subfigure[]{
  \includegraphics[width=4.7cm]{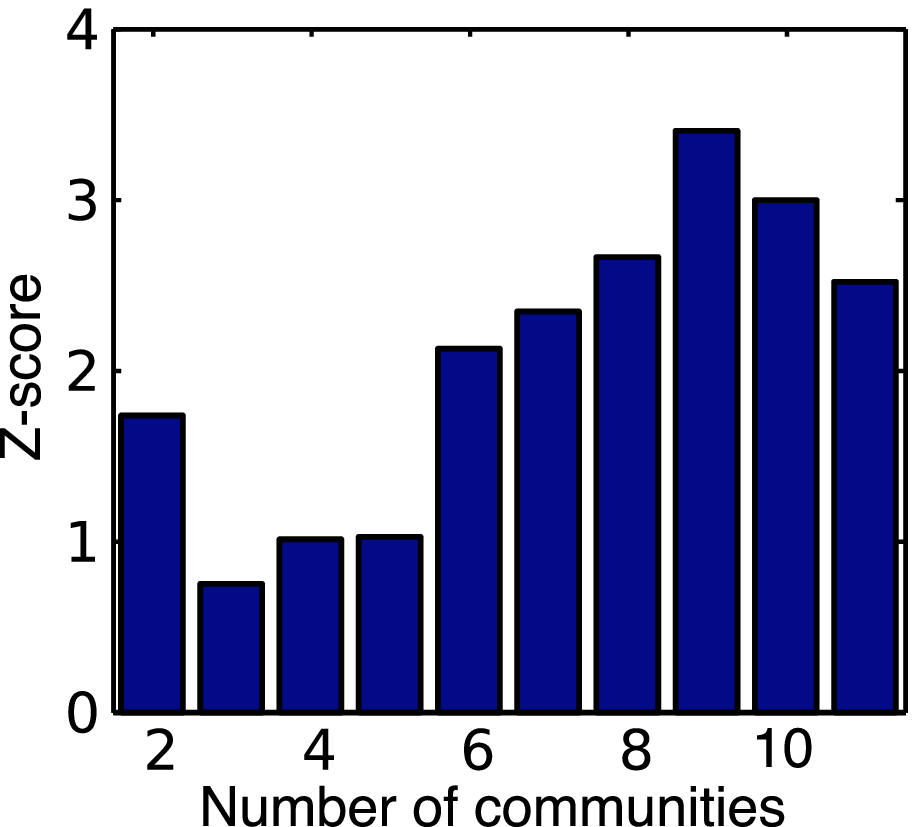}
  \label{fig:PfMTIPb}
  }
  \subfigure[]{
  \includegraphics[width=2.3cm]{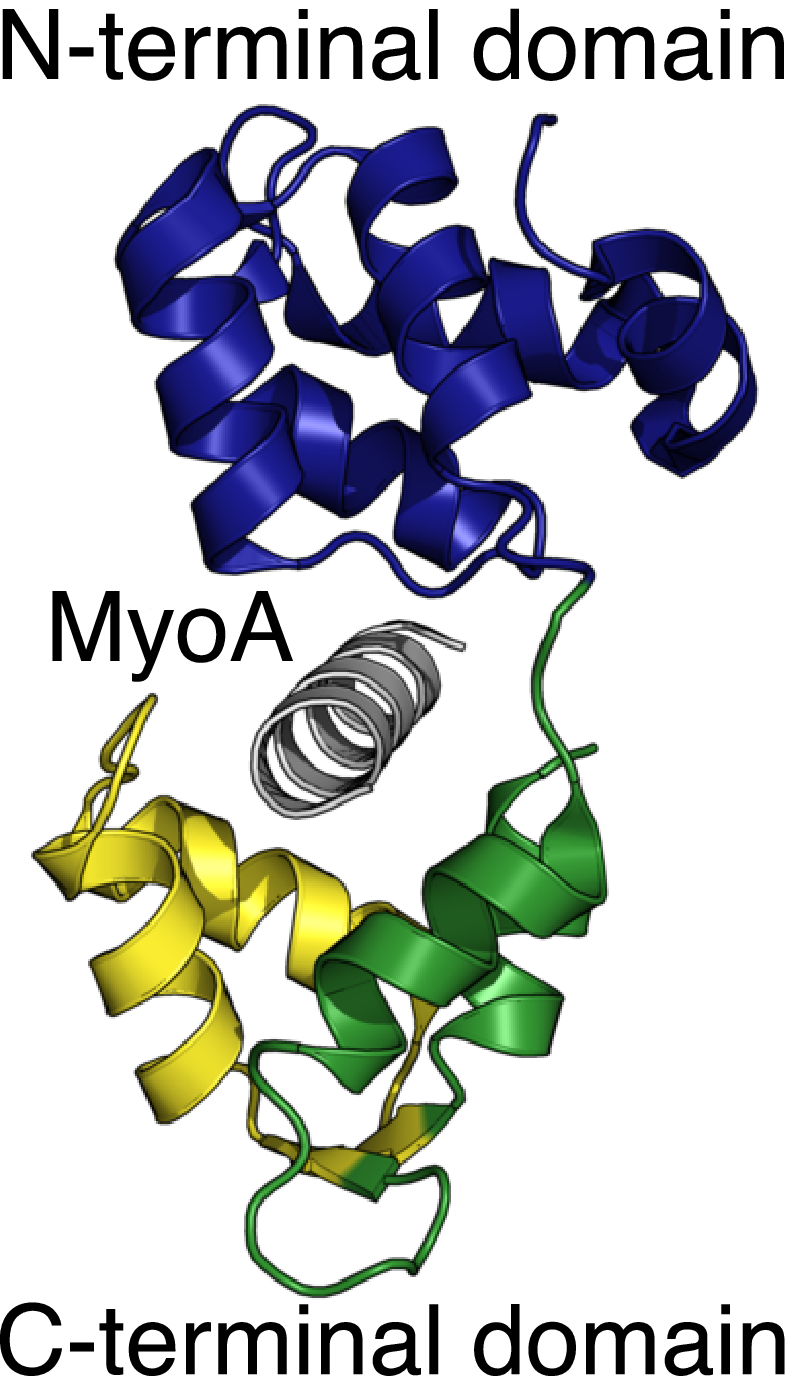}
  \label{fig:PfMTIPc}
  }\label{fig:PfMTIP}
\caption{(a) Multi-scale partitioning of PfMTIP/MyoA as a function of Markov time. The elements of the secondary structure are progressively grouped into larger communities as the Markov times evolves. Although in general our methodology does not pre-impose a hierarchical 
community structure, in this case the succession of community groupings is close to a strict hierarchy. Clusters kept for a long range of Markov times, such as the group of helices $\alpha$6 and $\alpha$7 are well-defined partitions. The identification of the rigid cluster, and of the functional domains leads to an increase in the robustness (z-score) of the partitions. (b) The comparison of the z-score per number of communities suggests that the partitions into 9 communities, where the rigid cluster is found, and the partition into two communities, where the N and C-terminal domains are identified, are significant. (c) Detection of the functional domains in the four-way partition of PfMTIP/MyoA. Protein structures drawn with PyMOL \cite{PyMOL}.}
\end{figure}

We have used the methodology introduced above to study the structure and dynamics of PfMTIP/MyoA[803-817], i.e., PfMTIP in complex with a peptide of the last 15 amino acids of the MyoA tail (PDB ID code 2QAC). Again, at small Markov times, we find partitions of high robustness corresponding to peptide bonds and amino acids (Fig.~\ref{fig:PfMTIPswap}). The relevant partitions at long Markov times are summarized in Fig.~\ref{fig:PfMTIPa}. Starting from the detection of the secondary structure, the different $\alpha$-helices and $\beta$ sheets are progressively incorporated in a quasi-hierarchical manner into bigger clusters as the Markov time increases. Some of the groupings lead to a marked increase in the robustness of the partition. This is the case for the first community to appear that incorporates two secondary structures: helices $\alpha$6 and $\alpha$7. This community is conserved across a broad range of Markov times, more than any other community of multiple elements of secondary structure. This suggests a strong dynamical linkage between these two $\alpha$-helices over an extended time scale of motion. This is in agreement with the PkMTIP crystal structure from Bosch~\emph{et al}~\cite{Bosh2006}, where these helices keep the same relative position in all three conformations observed, thereby suggesting the presence of a rigid cluster.

Another important community is the one formed by helices $\alpha$5 and $\alpha$8, which also leads to an increase in the robustness of the partition. Together with the $\alpha6-\alpha7$ cluster, they divide the C-terminal domain into two lobes that wrap around the MyoA peptide. The next rise in the z-score appears at Markov times around 500, at which point MTIP is divided into three domains: the two lobes of the C-terminal domain and the entire N-terminal domain (Fig.~\ref{fig:PfMTIPc}). The strong robustness of this particular partition reflects its significance for the functioning of the protein itself. This again supports hypotheses from Bosch~\emph{et al}~\cite{Bosch2007} concerning the closing mechanism of MTIP around MyoA, suggested to be in the form of a clamp, with the two lobes of the C-terminal domain wrapping around the MyoA tail, and the N-terminal domain fortifying the binding by bending towards the C-terminal domain to close the clamp. 

At long Markov times, the complex is partitioned into N- and C-terminal domains, with the MyoA peptide clustered with the C-terminal domain. This is also in agreement with results from K$_\mathrm{d}$ analyses~\cite{Jemima2010}, which suggest that the MyoA tail should be more tightly bound to the C-terminal than to the N-terminal domain. In addition, the fact that the two lobes of the C-terminal domain cluster at an earlier Markov time suggests that the wrapping of the C-terminal domain around the MyoA tail should take place more rapidly than the closing motion of the N-terminal domain.

\subsection{The partitions of PkMTIP suggest a stabilizing role of the MyoA tail and a strong similarity between PkMTIP and PfMTIP structures}

\begin{figure}[htp]
  \centering
  \subfigure[]{
  \includegraphics[width=8.5cm]{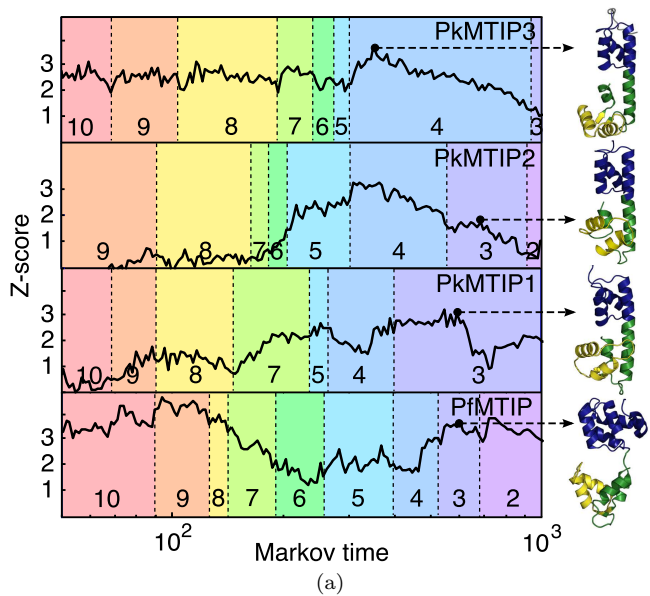}
  \label{fig:PkMTIPsa}
  }
  \subfigure[]{
  \includegraphics[width=7.5cm]{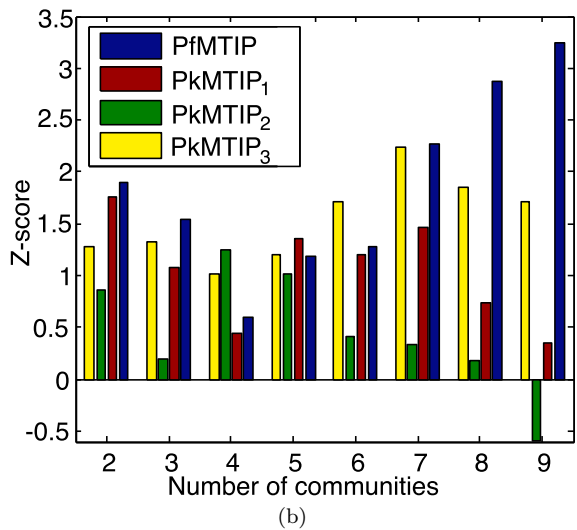}
  \label{fig:PkMTIPsb}
  }
  \label{fig:PkMTIPs}
\caption{(a) Robustness of the partitions of MTIP in different conformations as a function of the Markov time. The liganded conformations (PfMTIP and PkMTIP3) show better properties of robustness than the unliganded ones at the level of the secondary structures and of the functional domains, suggesting a stabilizing role of the binding with MyoA. Partitions are very similar between the three conformations, in particular for the functional domains, although the grouping of helices $\alpha$5 and $\alpha$8 only occurs at long Markov times for PkMTIP2. Protein structures drawn with PyMOL \cite{PyMOL}. (b) The z-score of the partitions with the same number of communities compared across different conformations of MTIP shows that the liganded forms have better defined partitions.}
\end{figure}

We have used our methodology to study the changes in the structural organization of MTIP induced by the presence of the MyoA peptide
by comparing unliganded (free) conformations of MTIP (PkMTIP1 and PkMTIP2) with `liganded' conformations of MTIP (PkMTIP3 and PfMTIP), which are obtained from MTIP-MyoA complexes by deleting the MyoA peptide together with all its interactions when generating the graph. Liganded conformations thus reflect the change of shape induced by the MyoA peptide with none of the direct constraints. Fig.~\ref{fig:PkMTIPsa}~and~\ref{fig:PkMTIPsb} show that the partitions for the liganded conformations obtained from the complexed forms are in general much more robust than the partitions of the two unliganded structures, especially at the level of the secondary structure (8, 9, and 10 communities) and of functional domains (3 communities). Such increase of the robustness of the partitions in the liganded conformations emerges naturally from the change in the spatial structure induced by the MyoA peptide.

Importantly, although the partitions differ significantly in their robustness and the Markov time of their predominance, they are very similar among the different conformations, especially between the two liganded forms. In particular, the important communities identified in section~\ref{sec:PfMTIP}, such as the $\alpha6-\alpha7$ cluster and the functional domains (Fig.~\ref{fig:PfMTIPc}), are also detected in all three conformations with high robustness (the only exception being the functional domains of PkMTIP2). The fact that the similar partitions are found in all structures suggests that the overall organization of the protein is not changed much between the conformations and corroborates the idea that the space of conformations that can be taken by a proteins is inscribed within its own structure~\cite{kern}. However, changes in the robustness and Markov lifetime of the partitions suggests that the secondary and tertiary structures get better structured upon binding with the MyoA tail since the corresponding partitions are better defined in this case. Note also that in comparing the liganded conformations, PfMTIP has more robust partitions than PkMTIP3, possibly a result of the stabilizing role of the N-terminal domain, which in PfMTIP also binds the peptide and closes the clamp. On the other hand, the unliganded form PkMTIP2 possesses the least robust partitions, in accordance with the hypothesis~\cite{Bosh2006} that it should be an intermediate conformation between the fully opened and fully closed forms.

\begin{figure}
 \centering
 \includegraphics[width=9cm]{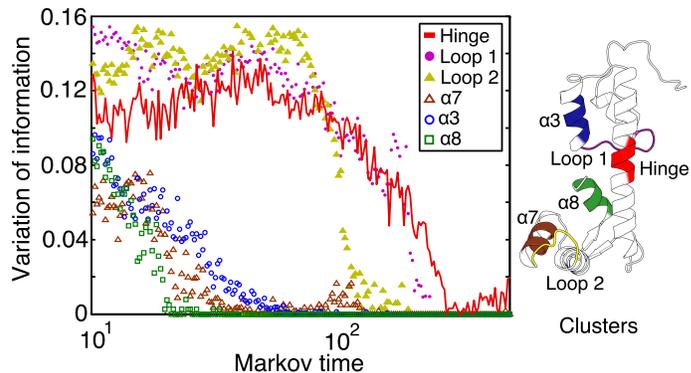}
 % PfMTIP_swap.eps: 0x0 pixel, 300dpi, 0.00x0.00 cm, bb=
 \caption{Variation of information of the partitions detected in different secondary structures. The variation of information of the central region of the central $\alpha$-helix (continuous red line) is very high and behaves similarly to other loop regions of the protein. This suggests that the algorithm effectively recognizes this region as a loop, despite its $\alpha$-helical secondary structure. Filled symbols correspond to loops,  empty symbols correspond to $\alpha$-helices and the continuous line corresponds to the hinge region. Protein structure drawn with PyMOL~\cite{PyMOL}.}
 \label{fig:robust_clusters}
\end{figure}

The similarity between the partitions of the complexed forms of PfMTIP and PkMTIP supports the expectation that their structural organization should not be very different. However, there is a current open debate in the literature regarding the difference in the hinge region between the N- and C-terminal domains, with PkMTIP presenting a long central $\alpha$-helix where PfMTIP only has a loop (Fig. \ref{fig:PkMTIPsa}). The classification of this central domain in PkMTIP as a helix is however controversial since the structure of PkMTIP was measured at a non-physiological pH and does not correspond to the structure observed in other MLCs~\cite{Bosh2006}. Interestingly, our partitioning consistently divides the central $\alpha$-helix of PkMTIP into two different communities at all Markov times. Furthermore, in the partitions, the separation between the two halves of this central $\alpha$-helix is constrained within the region that corresponds to the central loop in PfMTIP (from residues H135 to N140). This partitioning is thus consistent with the central $\alpha$-helix of PkMTIP being partly identified as a loop by the partitioning algorithm. To further support this observation, we
have carried out an analysis of the robustness of loops and $\alpha$-helices with the same number of nodes (50 atoms) across Markov times. Fig.~\ref{fig:robust_clusters} shows that the central region of the central $\alpha$-helix of PkMTIP has a robustness much lower than the typical $\alpha$-helix with a trend similar to that of loops. These results demonstrate the insights that our method can bring into the analysis of the structural organization of a protein beyond its pre-assigned secondary or tertiary structure.

\subsection{The analysis of residue sensitivity suggests six residues of particular importance for the structure and dynamics of the complex}

The last part of the analysis aims to identify residues in the MyoA tail that have a strong impact on the multi-scale organization of the protein complex and can therefore be considered to play a significant role in its structure and dynamics. This analysis does \textit{not} evaluate the influence of a mutation on the binding energy; rather, the expectation is that residues with a large influence on the structural organization of the protein will affect the global dynamics of the binding events. Indeed, hotspots are known experimentally to be related to the global mechanical properties of the protein such as flexibility and intrinsically disordered regions~\cite{Ma2001,Radivojac2007,kernAdK}. Furthermore, various computational methods have demonstrated the high influence of hotspots on large-scale attributes such as the distribution of conformations~\cite{Ming2006}, the network of cooperativity between residues measured in terms of coupled fluctuations~\cite{Liu2007}, or their propensity of being located at hinge sites measured by their mobility in the slow modes~\cite{Yang2005}.
To assess the connection between the effect on the binding energy and on the structural and dynamical features detected by our method, we have compared our computational results with the outcome of binding assays of MTIP with mutated MyoA tail peptides.

\begin{figure*}[htb]
  \centering
  \subfigure[]{
  \includegraphics[width=6.6cm]{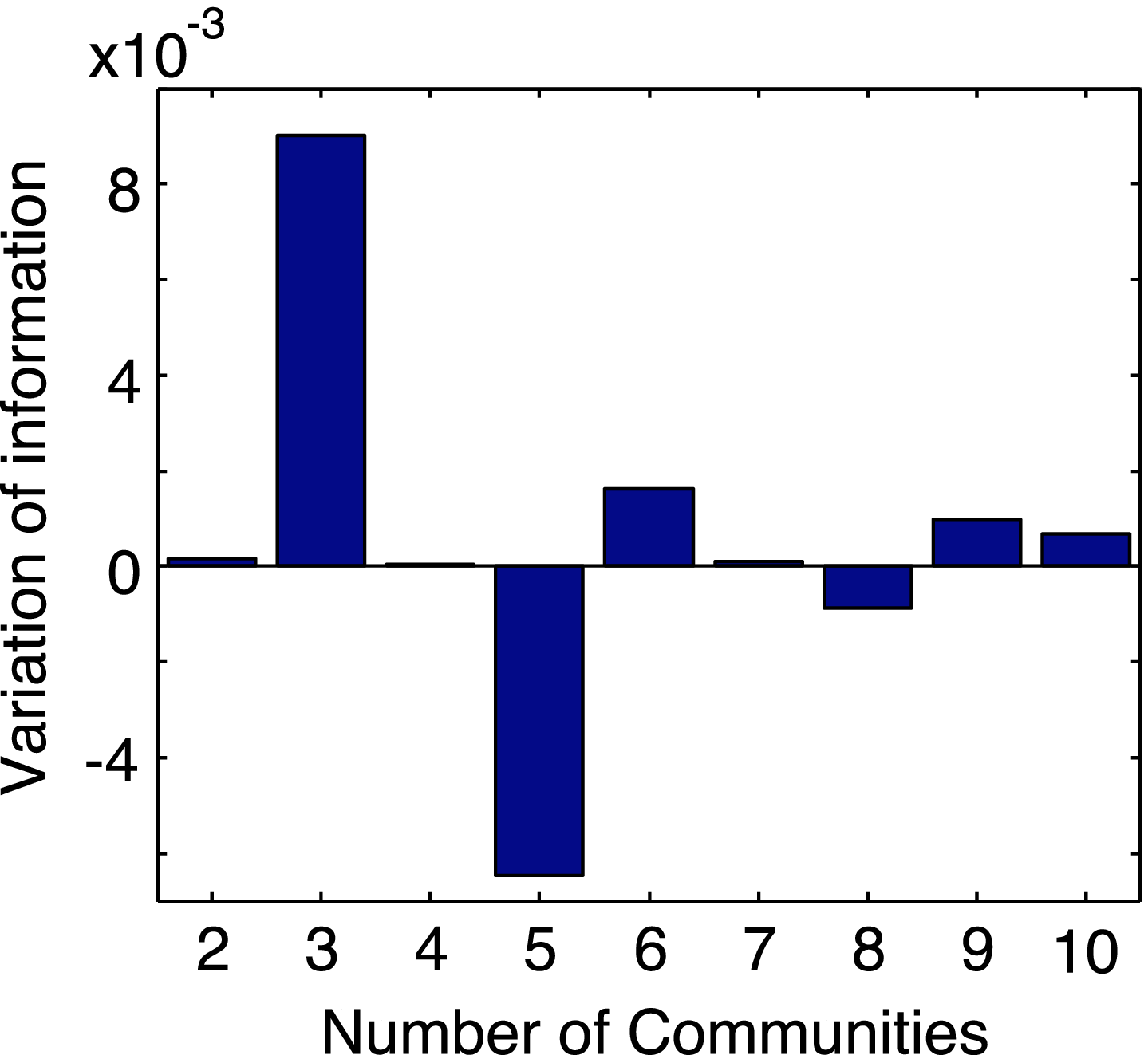}
  \label{fig:mutationsa}
  }
  \subfigure[]{
  \includegraphics[width=6.6cm]{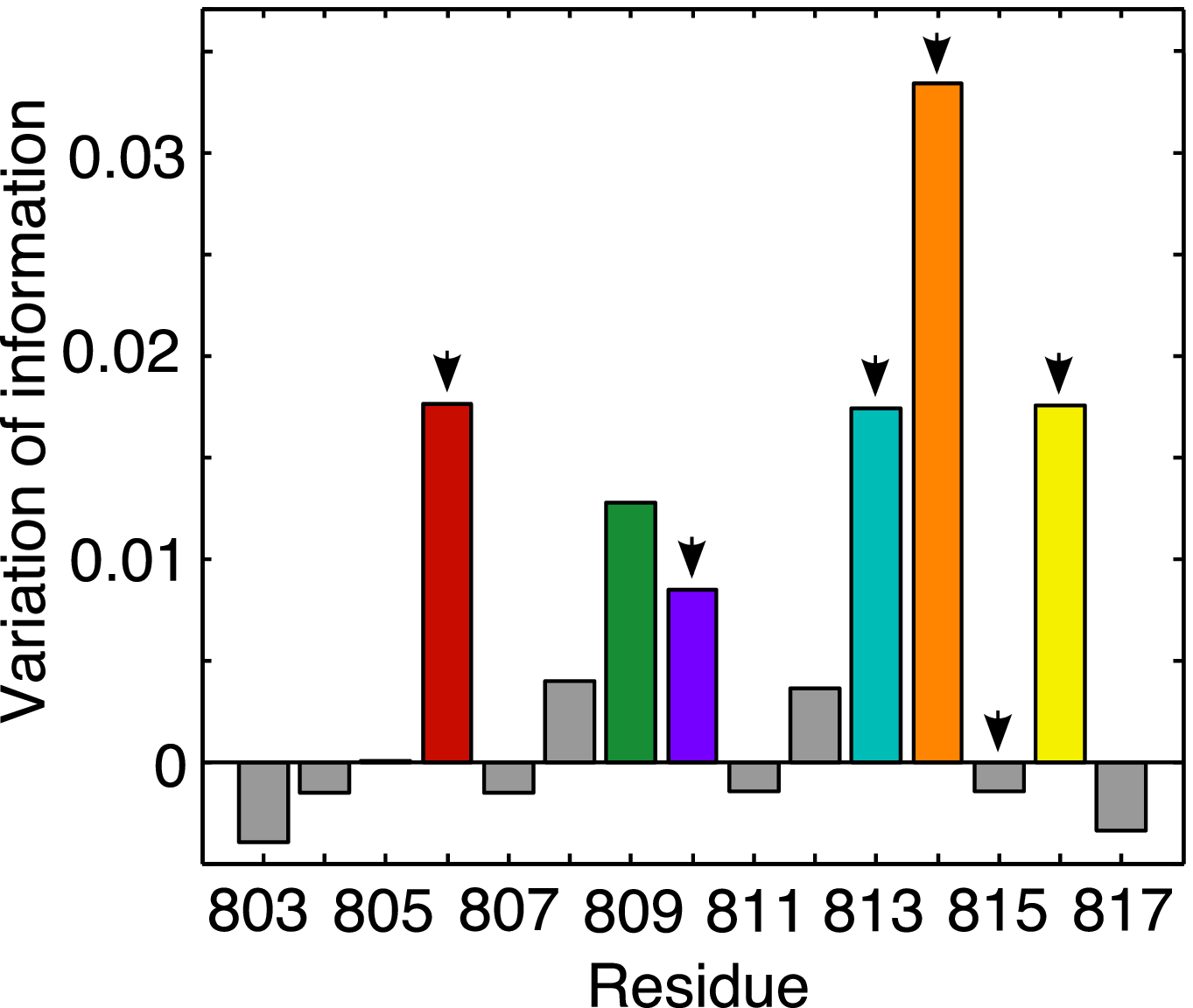}
  \label{fig:mutationsb}
  }
  \subfigure[]{
  \includegraphics[width=6.4cm]{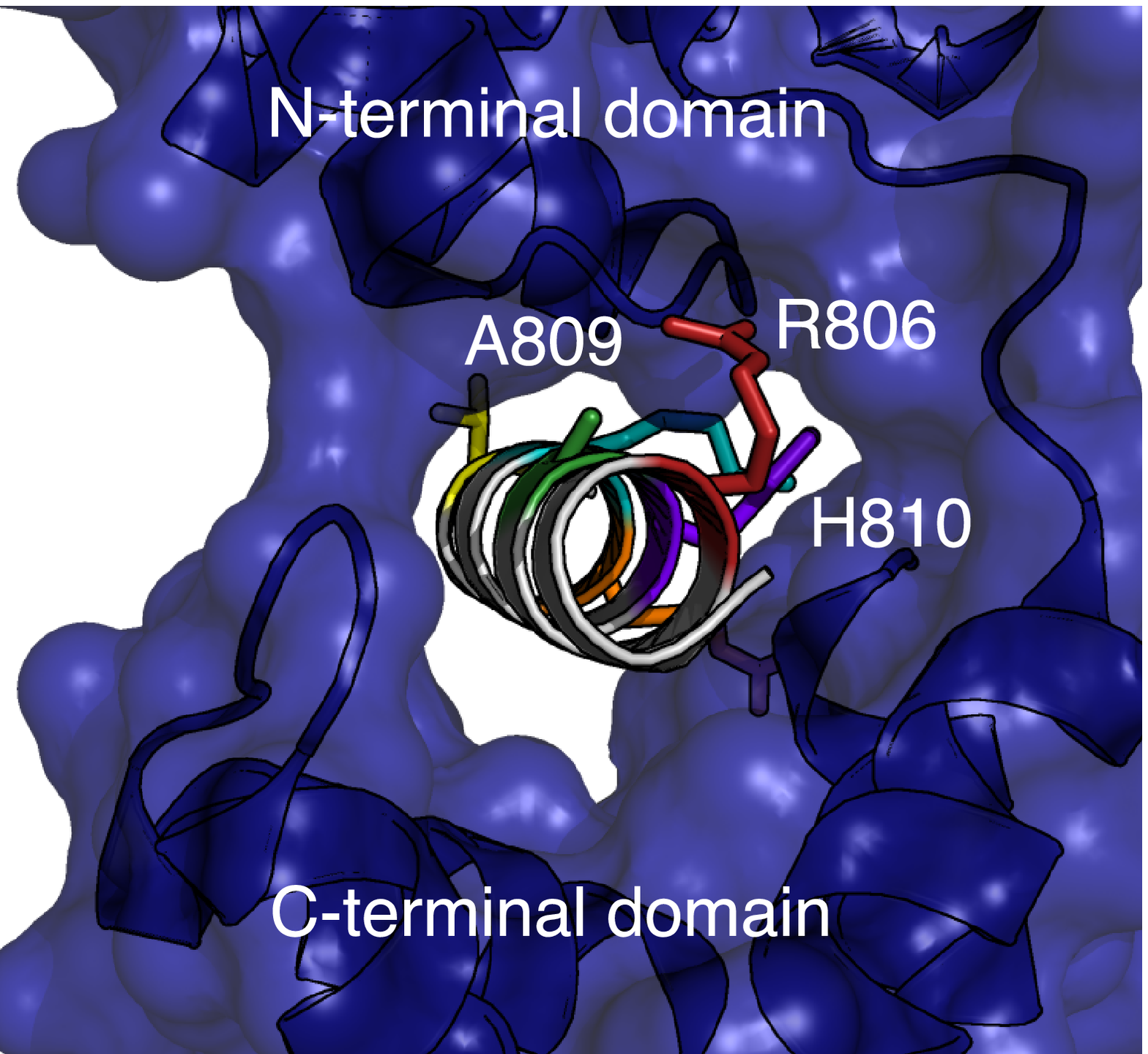}
  \label{fig:mutationsc}
  }
  \subfigure[]{
  \includegraphics[width=6.4cm]{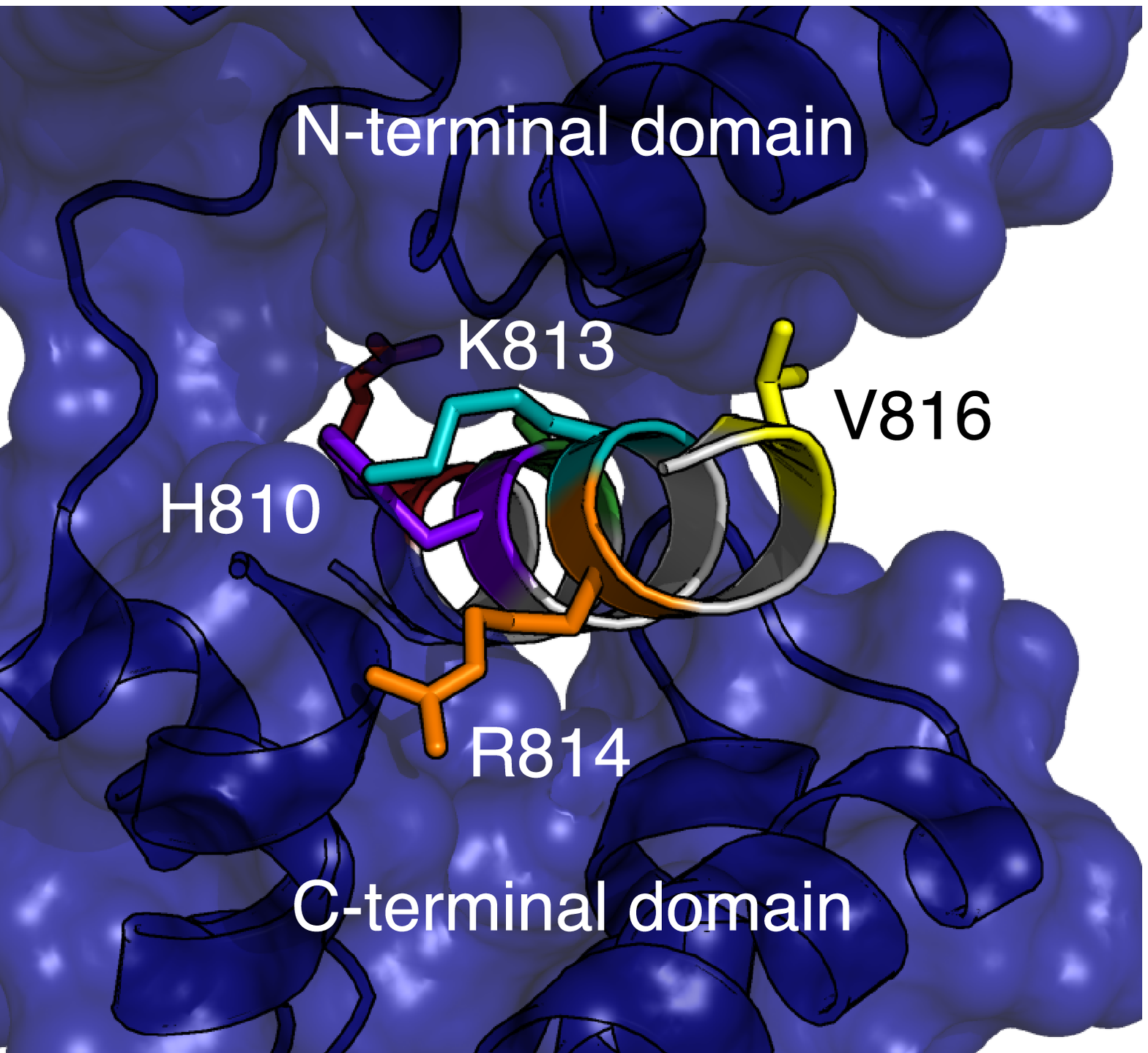}
  \label{fig:mutationsd}
  }
\caption{(a) The partition most influenced by computational mutagenesis is the one into 3 communities. (b) Residues R806, A809, H810, K813, R814, and V816 have the biggest influence on the three-way partition. There is a strong coincidence with the residues identified experimentally \cite{Jemima2010} to have a strong effect on the binding affinity (indicated by arrows). (c) and (d) Front and side view of the positions of the key residues found. Structures drawn with PyMOL \cite{PyMOL}.}\label{fig:mutations}
\end{figure*}

Our computational setup mimics the standard \textit{alanine scanning mutagenesis}: each residue is `mutated' in turn by removing from the graph all the edges corresponding to the weak interactions it makes with other residues. The mutated graph is then analyzed with our multi-scale methodology, and the partitions are compared with those of the original graph using the VI. For each mutation, we compute the VI between all the partitions found with the same number of communities from the original and mutated protein averaged over 10 different Louvain initial conditions and normalized by the average VI of the original graph. Using this scheme, partitions which are the most affected by a particular mutation will give a high value of the variation of information.

Fig. \ref{fig:mutationsa} shows that the partitions into 3 communities are the most affected by the mutations. This is not surprising since the 3-way partition is the first where the MyoA peptide is grouped with part of the MTIP molecule (Fig.~\ref{fig:PfMTIPa}). Consequently, the mutations essentially affect the strength of the association between the MyoA tail and the portion of MTIP that includes the hinge region and the helices $\alpha$5 and $\alpha$8 from the C-terminal domain. More specifically, the mutations that cause the largest changes in the 3-way partition are those in residues R806, A809, H810, K813, R814, and V816 (Fig.~\ref{fig:mutationsb}). These results are in accordance with experimental binding assays for MyoA peptides of different lengths \cite{Jemima2010}, crystallographic data and yeast two-hybrid experiments \cite{Bosch2007}. In particular, residues R806 and K813 have been observed to be essential for complex formation; H810 and R814 provide key contacts for tight binding; and V816 also improves the binding. On the other hand, our method does not single out a significant contribution of residue M815, which has been found to influence binding affinity. A possible explanation is that the importance of this residue might be related to effects that are not directly addressed by our method, such as intermediate states in the folding pathway
or modification of binding energies. Finally, our method finds one residue, A809, predicted to have an important effect on the multi-scale organization which has not been investigated experimentally to date.

\section{Conclusion and outlook}

We have presented the application of an efficient and computationally inexpensive method to extract information about the structural and dynamical organization of proteins across time and spatial scales starting bottom-up from the full atomistic information. The methodology is based on multi-scale graph partitioning methods that establish a series of increasingly coarser partitions that can reveal the structure of the graph. This paper introduces new graph theoretical tools, specifically the use of the robustness of partitions as a measure of their biological significance and the quantification
of robustness through the introduction of biochemically-motivated surrogate random graph models. 

Our analysis has uncovered important features of the MTIP/MyoA complex that agree well with experimental data. The rigid cluster formed by helices $\alpha$6 and $\alpha$7, as suggested by the crystal structures of PkMTIP \cite{Bosh2006}, was observed to form a well-defined community, conserved across a broad range of Markov times and associated with very robust partitions. The functional domains suggested by the analysis of crystal structures of different conformations across species~\cite{Bosh2006,Bosch2007} have been detected by the partitioning and also showed strong robustness and conservation across Markov times. The robustness analysis of the hinge region of PkMTIP confirms these similarities between species and therefore suggests that their dynamical behavior should be similar. Furthermore, it supports the hypothesis~\cite{Bosch2007} that the reported differences between PkMTIP and PfMTIP in the the hinge region could result from the particularities of the crystallization. Finally, a computational tool for mutational analysis was introduced and used to identify five out of the six residues known from binding assays \cite{Jemima2010} to have a strong influence on the binding of MyoA. It also suggested one additional residue, A809, which has not yet been investigated experimentally, to be particularly important.

Together, these results provide a better understanding of the possible dynamical behavior of MTIP and other myosin light chains. The broad agreement with a variety of experimental results underlines the intrinsic interest of methods that study the multi-scale organization of proteins. Our method includes atomistic detail encoded in a graph representation and allows to extract information about the global organization of the structure and dynamics of the protein, but also about how individual residues can affect them. Future work will include the experimental verification of these predictions, such as binding assays with alanine mutations of some of the key residues identified. A deeper study of the role of each residue of the MyoA tail could also be carried out by computationally analyzing their effect on individual communities, rather than on the whole partitioning. On the theoretical side, although the two random graph models proposed here were shown to generate relevant null hypotheses, other forms of randomization such as geometric random graphs could also be tested in the future. Finally, using the communities detected by our method to coarse-grain molecular dynamics simulations could provide an efficient method to get insight into the folding and closing pathway.

\begin{acknowledgments}
AD was supported through a PhD Studentship Award from the British Heart Foundation Centre of Research Excellence from Imperial College London and (grant number RE/08/002) a Wallonie-Bruxelles International Award. EWT acknowledges support through a David Phillips Fellowship from the Biotechnology and Biological Sciences Research Council. We thank Stefano Meliga, Renaud Lambiotte, Jean-Charles Delvenne, Joao Costa and YunWilliam Yu for helpful discussions.
\end{acknowledgments}

%\appendix 
%\setcounter{section}{1} 
%\section*{Appendix}

%\section*{Abbreviations list} 

%\newpage
%\section*{References}
\bibliographystyle{iopart-num}
\bibliography{delm0419.bib}

%\section*{Figure legends}

%\section*{Tables}

%\section*{Glossary}

\end{document}